\documentclass[10pt,twocolumn,prl,aps,amssymb,amsmath,tightenlines]{revtex4}
\usepackage{dsfont}
\usepackage{graphicx}
\usepackage{amssymb,amsfonts}

\newcommand{\half}{\mbox{$\textstyle\frac{1}{2}$}}
\newcommand{\re}{\mbox{$\rm e$}}
\newcommand{\ri}{\mbox{$\rm i$}}
\newcommand{\rd}{\mbox{$\rm d$}}

\begin{document}
\title{Making sense of the divergent series for reconstructing a Hamiltonian \\
from its eigenstates and eigenvalues}
\author{Carl M. Bender$^1$, Dorje C. Brody$^2$, and Matthew F. Parry$^3$}

\affiliation{$^1$Department of Physics, Washington University, St. Louis, MO
63130, USA\\ 
$^2$Department of Mathematics, University of Surrey, Guildford GU2 7XH, UK\\ 
$^3$Department of Mathematics and Statistics, University of Otago, Dunedin 9054,
New Zealand}

\date{\today}

\begin{abstract}
In quantum mechanics the eigenstates of the Hamiltonian form a complete basis. However, physicists conventionally express completeness as a formal sum over the eigenstates, and this sum is typically a divergent series if the Hilbert space is infinite dimensional. Furthermore, while the Hamiltonian can be reconstructed formally as a sum over its eigenvalues and eigenstates, this series is typically even more divergent. For the simple cases of the square-well and the harmonic-oscillator potentials this paper explains how to use the elementary procedure of Euler summation to sum these divergent series and thereby to make sense of the formal statement of the completeness of the formal sum that represents the reconstruction of the Hamiltonian.
\end{abstract}

\maketitle
In physics courses quick formal arguments are often presented to reach correct
conclusions that may be quite difficult to justify at a mathematical level. An
example of such a formal argument is the derivation of the spectral
decomposition of a Hamiltonian (or of any other observable). If $\{|\phi_n
\rangle\}$ are the normalized eigenvectors of a Hermitian Hamiltonian ${\hat H}$
and $\{E_n\}$ are the corresponding eigenvalues, then the Hamiltonian is usually
represented (reconstructed) formally as the sum
\begin{eqnarray}
{\hat H}=\sum_n E_n|\phi_n\rangle\langle\phi_n|. 
\label{E1}
\end{eqnarray}
There is no problem with (\ref{E1}) if the Hilbert space is {\it
finite-dimensional}; that is, if ${\hat H}$ is an $N\times N$ matrix. In such a
case $\hat H$ is given explicitly by the {\it finite} sum $\sum_{n=1}^N E_n|
\phi_n \rangle\langle\phi_n|$. For $N=2$ or $N=3$, an explicit calculation of
this sum is a useful exercise in an introductory quantum mechanics course.

However, a thoughtful student might ask, For a square-well potential on the
interval $[0,\pi]$ the normalized eigenfunctions of the Hamiltonian are 
$\phi_n(x)=\sqrt{2/\pi}\sin(nx)$ and the corresponding eigenvalues are $E_n=
\half n^2$. How can I evaluate the {\it infinite} sum 
\begin{eqnarray}
\sum_{n=1}^\infty E_n\phi_n(x)\phi_n(y)=\frac{1}{\pi}\sum_{n=1}^\infty 
n^2\sin(nx)\sin(ny)
\label{E2} 
\end{eqnarray}
to obtain the coordinate-space representation $H(x,y)=\langle x|{\hat H}|y
\rangle=-\half\delta''(x-y)$ of the Hamiltonian ${\hat H}=\half{\hat p}^2$?
The problem here is that the series (\ref{E2}) diverges. A mathematically
precise treatment for the reconstruction of the Hamiltonian may be found in
mathematical literature \cite{rs} but not in standard textbooks on quantum
mechanics \cite{texts}. Furthermore, the formal series  
\begin{eqnarray}
\delta(x-y)=\sum_n\langle x|n\rangle\langle n|y\rangle=\sum_n 
{\bar\phi}_n(x)\phi_n(y),
\label{E3} 
\end{eqnarray}
which is conventionally used to express the completeness of the eigenstates, is
also divergent for the square-well even though it diverges less rapidly than the 
series (\ref{E2}). 

The series analogous to (\ref{E2}) for reconstructing the Hamiltonian of a
quantum harmonic oscillator is also divergent. The $n$th term in this series
grows for large $n$ like $\sqrt{n}$ multiplied by an oscillatory term. This sum
is better behaved than that in (\ref{E2}), but it still diverges. 

In what sense can we interpret the divergent series (\ref{E2}) as the
Hamiltonian for the square-well potential? Is it possible to sum the divergent
series in (\ref{E3})? In this paper we propose to use the procedure of {\it
Euler summation} as a simple and sensible way to regulate these divergent 
series and thereby to obtain closed-form coordinate-space representations 
$H(x,y)$ of the Hamiltonian and $\delta(x-y)$ of the identity operator. 

Euler summation is easy to perform: Suppose that the series 
$\sum_{n=0}^\infty a_n$ diverges
but that $f(t)=\sum_{n=0}^\infty t^n a_n$ converges for $|t|<1$. The Euler sum $E$ of the
series $\sum_{n=0}^\infty a_n$ is then defined as the limit
$$E\equiv\lim_{t\to1^{-}}f(t)$$
{\it if this limit exists} \cite{euler}.

To illustrate Euler summation and demonstrate its effectiveness we use it to
evaluate a sum that represents the Riemann zeta function $\zeta(s)$. The zeta
function is conventionally defined in terms of the series 
\begin{eqnarray}
\zeta(s)=\sum_{n=1}^\infty\frac{1}{n^s}, 
\label{E4}
\end{eqnarray}
which converges for ${\rm Re}(s)>1$, but diverges for ${\rm Re}(s)\leq1$. 
Analytic continuation may be used to construct a complex integral representation
of $\zeta(s)$ that is valid for all values of $s\neq1$:
$$\zeta(s)=\frac{\Gamma(1-s)}{2\pi i}\oint_C \rd t\frac{t^{s-1}}{\re^{-t}-1},$$
where $C$ is a Hankel contour that encircles the negative-$t$ axis in the
positive direction. One can use this integral representation to calculate $\zeta
(s)$ at values of $s$ for which the sum (\ref{E4}) does not converge. For
example, when $s=0$ and when $s=-1$ we get
$$\zeta(0)=-\frac{1}{2},\qquad\zeta(-1)=-\frac{1}{12}.$$ 
These results are remarkable because they suggest in some formal sense that the 
divergent sum $1+1+1+\cdots$ has the value $-\frac{1}{2}$ and that the
divergent sum $1+2+3+\cdots$ has the value $-\frac{1}{12}$. 

Summation of divergent series such as these is not just a formal mathematical
procedure. It may be used to solve physical problems involving divergent sums
over vibrational modes. For example, calculating the Casimir force, which has
been verified and measured in laboratory experiments, requires that divergent
sums over physical modes be evaluated \cite{casimir}.

What happens if we attempt to use Euler summation to evaluate the series 
$1+1+1+\cdots$? The series for the Euler function $f(t)=1+t+t^2+\cdots$
converges for $|t|<1$ and we get 
$$f(t)=\frac{1}{1-t}\quad(|t|<1).$$
However, $\lim_{t\to1}f(t)$ does not exist. Thus, Euler summation is not
powerful enough to assign a value to the sum of the divergent series $1+1+1+
\cdots$. Euler summation also fails to assign a value to the sum of the
divergent series $1+2+3+\cdots$.

However, there is a clever way to use Euler summation to evaluate $\zeta(0)$: We
rewrite the series representation (\ref{E4}) for $\zeta(s)$ as an {\it
alternating} series by subtracting the even-$n$ terms from the odd-$n$ terms in
(\ref{E4}):
\begin{eqnarray}
\zeta(s)=\frac{1}{1-2^{1-s}}\sum_{n=1}^\infty\frac{(-1)^{n+1}}{n^s}.
\label{E5}
\end{eqnarray}
This alternating series converges for ${\rm Re}(s)>0$ but diverges for ${\rm Re}
(s)\leq0$. Let us use Euler summation to evaluate the series obtained by setting
$s=0$, which is $-1+1-1+1-\cdots$. For this divergent series $f(t)=-1/(1+t)$,
and thus the Euler sum $E$ of the series is $f(1)=-\frac{1}{2}$, which is the
correct value for $\zeta(0)$. 

Similarly, if we set $s=-1$ in (\ref{E5}), we obtain the divergent series $-
\frac{1}{3}+\frac{2}{3}-\frac{3}{3}+\frac{4}{3}-\cdots$. For this case 
$$f(t)=-\frac{1}{3}+\frac{2}{3}t-\frac{3}{3}t^2+\frac{4}{3}t^3-\cdots=-\frac{1}
{3(1+t)^2}.$$
Thus, Euler summation gives $\zeta(-1)=f(1)=-\frac{1}{12}$. Evidently, Euler
summation is quite impressive; it implicitly performs an analytic continuation
of the sums in (\ref{E4}) and (\ref{E5}) into the complex plane without the
appearance of complex numbers! 

Having demonstrated the power of Euler summation we now use it to regulate and
evaluate divergent quantum-mechanical sums like that in (\ref{E2}) and thus
make sense of this formal series. A rigorous discussion of completeness requires
advanced mathematical techniques used in Hilbert-space theory, but here we show
how to perform the sum in (\ref{E1}) explicitly for the special case of a
square-well potential and we repeat the process for a harmonic-oscillator
potential. We use only elementary techniques that are within reach of 
undergraduate physics students.

\vspace{0.2cm} 

\noindent {\bf Square-well potential}. A unit-mass particle trapped in an
infinite square-well potential on the interval $[0,\pi]$ is described by the 
coordinate-space Hamiltonian
$\hbar=1$) 
\begin{eqnarray} 
{\hat H}=-\frac{1}{2}\frac{\rd^2}{\rd x^2}
\nonumber
\end{eqnarray} 
(in units such that Planck's constant is unity). The normalized solutions to the
time-independent Schr\"odinger equation that satisfy vanishing boundary
conditions at $x=0$ and $x=\pi$ are
\begin{eqnarray}
\phi_n(x)=\sqrt{\frac{2}{\pi}}\sin(nx)\quad{\rm and}\quad E_n=\half n^2,
\nonumber
\end{eqnarray}
and thus from (\ref{E1}) we obtain the divergent series (\ref{E2}). 

A naive summation of (\ref{E2}) that one might encounter in an elementary
quantum-mechanics course consists of using the completeness condition
\begin{eqnarray}
\frac{2}{\pi}\sum_{n=1}^\infty\sin(nx)\sin(ny)=\delta(x-y) 
\label{E6} 
\end{eqnarray}
to verify the coordinate-space representation of the square-well Hamiltonian.
The argument begins by replacing the factor of $n^2$ in the formal series
\begin{eqnarray}
H(x,y)=\frac{1}{\pi}\sum_{n=1}^\infty n^2\sin(n x)\sin(n y)
\label{eq:7}
\end{eqnarray} 
with a second derivative acting on the full series:
$$H(x,y)=-\frac{1}{2}\frac{\partial^2}{\partial y^2}\left(\frac{2}{\pi}\sum_{n=
1}^\infty\sin(n x)\sin(n y)\right).$$
Then (\ref{E6}) is used to write
$$H(x,y)=-\frac{1}{2}\frac{\partial^2}{\partial y^2}\,\delta(x-y).$$
The problem with this argument is that the interchange of differentiation and
summation is justified only if the sum in (\ref{E6}) is absolutely and uniformly
convergent and this is not so because the formal series (\ref{E6})
diverges.

Let us now use Euler summation to make sense of the formal
statement of completeness in (\ref{E6}). We do so at a freshman-calculus level
and without appealing to advanced theorems about Fourier series. The usual
approach in Fourier analysis \cite{wiener} relies on regulating the infinite sum
(\ref{E6}) by replacing it with an $N$-term finite sum and then taking the limit
$N\to\infty$ by using the Riemann-Lebesgue lemma to argue that this sum
converges to a delta function. Instead, here we regulate the infinite sum
(\ref{E6}) by inserting the geometrical convergence (Euler) factor $t^n$:
\begin{eqnarray}
K(x,y,t) &=& \frac{2}{\pi}\sum_{n=1}^{\infty}t^n\sin(n x)\sin(n y)\nonumber\\
&& \hspace{-1.5cm}=\frac{1}{\pi}\sum_{n=1}^{\infty}t^n\left\{\cos[n(x-y)]- 
\cos[n(x+y)]\right\}.
\label{E8}
\end{eqnarray}
This sum converges absolutely and uniformly for $|t|<1$. We then use the
exponential form of the cosine function, $\cos(nz)=\half(\exp({\rm i}nz)+\exp(-{
\rm i}n z))$, and sum the infinite geometric series:
$$\sum_{n=1}^{\infty}t^n\cos(nz)=\frac{1}{2-2t\,\re^{{\rm i}z}}
+\frac{1}{2-2t\,\re^{-{\rm i}z}}-1.$$
Thus, we find that
$$K(x,y,t)=D(x-y,t)-D(x+y,t),$$ 
where
\begin{equation}
D(z,t)=\frac{1-t\cos z}{\pi(1-2t\cos z+t^2)}.
\label{E9}
\end{equation}

It is easy to show that $D(z,t)$ has the following properties: If $z=n\pi$
with $n$ even, then $D(z,t)=[\pi(1-t)]^{-1}$, whereas for other values of $z$,
$D(z,t)\to(2\pi)^{-1}$ as $t\to1$. Therefore, as $t\to1$ and for $0<x,y<\pi$,
the value of
$K(x,y,t)$ becomes infinite along the line $x=y$ and is zero otherwise. These
properties suggest that $K(x,y,t)\to\delta(x-y)$ as $t\to1$ and we must show
that this is so. In general, a parametric family of functions $\{f_t(x)\}$
having the property that $f_t(x)\to\delta(x)$ as $t\to t_0$ is said to be {\it
delta convergent}. (For a precise definition of a delta-convergent function
see Ref.~\cite{gelfand}.) In practice, to show that $\{f_t(x)\}$ is delta
convergent, it suffices to show that for any $a<0<b$ 
$$\lim_{t\to t_0}\int_a^b f_t(x)\rd x=1$$ 
and that if $a<b<0$ or if $0<a<b$, the limit $t\to t_0$ of the integral
vanishes. Thus, to establish completeness we integrate $K(x,y,t)$ in the
variable $y$
and then take the limit $t\to1^-$. This calculation is straightforward but
lengthy, so we have relegated it to the Appendix. It is shown there that
$K(x,y,t)$ in (\ref{E8}) is a delta-convergent series. This verifies the
completeness condition (\ref{E6}). 

Our next task is to use Euler summation to make sense of the divergent series
(\ref{eq:7}) for the reconstruction of the Hamiltonian. As before, we introduce
the convergence factor $t^n$ into (\ref{eq:7}):
$$H(x,y,t)=\frac{1}{\pi}\sum_{n=1}^\infty n^2 t^n \sin(n x)\sin(n y),$$
where $|t|<1$. The Euler-summation factor $t^n$ regulates the divergent
coordinate-space representation (\ref{eq:7}) for the Hamiltonian; the regulated
series $H(x,y,t)$ is absolutely and uniformly convergent. Therefore,
term-by-term differentiation of $H(x,y,t)$ with respect to the variables $x$,
$y$, and $t$ can be performed. 
We differentiate term-by-term with respect to $y$ to obtain
$$H(x,y,t)=-\frac{1}{2}\frac{\partial^2}{\partial y^2}K(x,y,t).$$
Therefore, in the limit $t\to1^-$, $H(x,y,t)$ converges to the coordinate-space
representation $H(x,y)=-\half\delta''(x-y)$ of the 
Hamiltonian for the square-well potential. 

\vspace{0.2cm} 

\noindent {\bf Quantum harmonic oscillator}. The coordinate-space Hamiltonian
that describes a particle of unit mass trapped in a harmonic potential is
\begin{eqnarray}
{\hat H}=-\frac{1}{2}\frac{\rd^2}{\rd x^2}+\frac{1}{2}x^2. 
\label{E10}
\end{eqnarray}
The stationary states and associated eigenvalues are 
$$\phi_n(x)=\frac{1}{\sqrt{2^n n!\surd{\pi}}}\re^{-\frac{1}{2}x^2}\!H_n(x)
\quad\!{\rm and}\!\quad E_n=n+\half.$$
Here, $H_n(x)$ denotes the $n$th Hermite polynomial, which can be obtained from
the standard identity 
\begin{eqnarray}
H_n(x)=(-1)^n\re^{x^2}\frac{\rd^n}{\rd x^n}\,\re^{-x^2}. 
\label{E11}
\end{eqnarray} 
Thus, from (\ref{E1}) the formal coordinate-space representation for the
Hamiltonian takes the form of the infinite sum
$$H(x,y)=\frac{1}{2\sqrt{\pi}}\,\re^{-\frac{1}{2}(x^2+y^2)}\sum_{n=0}^\infty 
\frac{2n+1}{2^n n!}H_n(x)\,H_n(y).$$ 
While this series is divergent, we show below how to use Euler summation to sum
the series to obtain the operator ${\hat H}$ in (\ref{E10}). 

As in the case of the square-well potential, the procedure is first to use Euler
summation to sum the formal series that represents the completeness of the
harmonic-oscillator eigenstates. To do so, we need a formula known as the {\it
Mehler generating function}:
$$K(x,y,t)=\sum_{n=0}^\infty t^n\phi_n(x)\phi_n(y).$$

Perhaps the simplest derivation of the Mehler formula for $K(x,y,t)$ 
was given by Hardy in his
lectures on orthogonal polynomials delivered in the Lent Term, 1933
(see \cite{watson}). 
We reproduce Hardy's derivation of the Mehler formula 
here for completeness. To
begin, we remark that the Fourier transform of a Gaussian is a Gaussian: 
\begin{equation}
\re^{-x^2}=\frac{1}{\sqrt{\pi}}\int_{-\infty}^\infty\re^{-u^2+2{\rm i}xu}\rd u.
\label{E12} 
\end{equation}
Differentiating (\ref{E12}) $n$ times in $x$ and using (\ref{E11}), we find that
$$H_n(x)=\frac{(-2\ri)^n\re^{x^2}}{\sqrt{\pi}}\int_{-\infty}^\infty 
u^n\re^{-u^2+2{\rm i}xu}\rd u.$$
It follows that 
\begin{eqnarray}
&& \sum_{n=0}^\infty t^n\phi_n(x)\phi_n(y)=\pi^{-\frac{3}{2}}\re^{\frac{1}{2}
(x^2+y^2)}\nonumber\\
&&\times\sum_{n=0}^\infty\int\limits_{-\infty}^\infty\!\!\int\limits_{-\infty
}^\infty\frac{(-2tuv)^n}{n!}\re^{-u^2-v^2+2{\rm i}(xu+yv)}\rd u\rd v.\nonumber 
\end{eqnarray}
When $|t|<1$ we can interchange the order of summation and integration because 
the series is uniformly and absolutely convergent and we get the Mehler formula
\begin{eqnarray}
&& K(x,y,t)=\pi^{-\frac{3}{2}}\re^{\frac{1}{2}(x^2+y^2)}\nonumber\\
&& \hspace{0.80cm}\times\int\limits_{-\infty}^\infty\!\!\int\limits_{-\infty
}^\infty \re^{-u^2-2tuv-v^2+2{\rm i}(xu+yv)}\rd u\rd v\nonumber\\
&& =\pi^{-1} \re^{\frac{1}{2}(x^2-y^2)}\int\limits_{-\infty}^\infty\re^{-(1-t^2)
u^2+2{\rm i}(x-yt)u}\rd u \nonumber\\
&& =\frac{1}{\sqrt{\pi(1-t^2)}}\exp\left[\frac{x^2-y^2}{2}-\frac{(x-yt)^2}{1-t^2}
\right].
\label{E13}
\end{eqnarray}

The last expression in (\ref{E13}) is not explicitly symmetric in the variables
$x$ and $y$, but it can be symmetrized by means of an elementary manipulation 
(see Ref.~\cite{wiener}, \S8): 
\begin{eqnarray}
&& \frac{x^2-y^2}{2}-\frac{(x-yt)^2}{1-t^2}\nonumber\\
&& \qquad\quad=-\frac{1-t}{1+t}\frac{(x+y)^2}{4}-\frac{1+t}{1-t}\frac{(x-y)^2}
{4}. 
\label{E14} 
\end{eqnarray}
Using (\ref{E14}), we observe that in the limit $t\to1^-$ the expression in
(\ref{E13}) assumes a Gaussian form with vanishing standard deviation (a Dirac
delta function). Indeed, we can easily integrate a Gaussian density explicitly 
to show that in the limit of vanishing standard deviation a Gaussian
density is delta convergent (see Ref~\cite{gelfand}, \S2.5). 
Hence, we deduce the completeness condition for the
eigenstates $\{\phi_n(x)\}$ of the quantum harmonic oscillator: 
$$\delta(x-y)=\frac{1}{\sqrt{\pi}}\,\re^{-\frac{1}{2}(x^2+y^2)}
\sum_{n=0}^\infty\frac{1}{2^n n!}H_n(x)\,H_n(y).$$ 
 
To go from this statement of completeness to the series reconstruction of the
Hamiltonian we again observe that differentiation of $K(x,y,t)$ in $t$ or in $y$
can be performed under the summation because the sum for $K(x,y,t)$ is uniformly
and absolutely convergent for $|t|<1$. Exploiting the relation
$$\left(t\frac{\rd}{\rd t}+\frac{1}{2}\right)t^n=\left(n+\frac{1}{2}\right)t^n,
$$ 
we show that
\begin{eqnarray}
H(x,y) &=& \lim_{t\to1}\left(t\frac{\rd}{\rd t}+\frac{1}{2}\right)\sum_{n=0
}^\infty t^n\phi_n(x)\phi_n(y)\nonumber\\
&=& \lim_{t\to1}\left(-\frac{1}{2}\frac{\partial^2}{\partial y^2}+\frac{1}{2}
y^2\right)\sum_{n=0}^\infty t^n\phi_n(x)\phi_n(y)\nonumber\\
&=& \left(-\frac{1}{2}\frac{\partial^2}{\partial y^2}+\frac{1}{2}y^2\right)
\delta(x-y).\label{eq:15} 
\end{eqnarray} 
Note that in going from the first to the second line above we use the identity
\begin{eqnarray}
&& \left( t\frac{\rd}{\rd t}+\frac{1}{2}\right) \frac{1}{\sqrt{\pi(1-t^2)}}\,\re^{
\frac{x^2-y^2}{2}-\frac{(x-yt)^2}{1-t^2}}\nonumber\\
&& \qquad =\left(-\frac{1}{2}\frac{\partial^2}{\partial y^2}+\frac{1}{2}y^2
\right)\frac{1}{\sqrt{\pi(1-t^2)}}\,\re^{\frac{x^2-y^2}{2}-\frac{(x-yt)^2}{1-t^2}}.
\nonumber
\end{eqnarray} 
Alternatively, making use of the eigenvalue equation 
$$ \left(-\frac{1}{2}\frac{\partial^2}{\partial y^2}+\frac{1}{2}
y^2\right) \phi_n(y) = \left(n+\frac{1}{2}\right) \phi_n(y) $$ 
satisfied by the harmonic oscillator eigenfunctions 
in the second line of (\ref{eq:15}) we arrive at the same conclusion more 
expediently. This completes the analysis for the quantum-harmonic-oscillator 
Hamiltonian.

\vspace{0.2cm} 

\noindent {\bf Summary}. We have used Euler summation to make sense of the
formal divergent series that express the completeness of the square-well and the
harmonic-oscillator eigenstates. We then showed how to interpret the
divergent series representing the coordinate-space reconstruction of the
corresponding Hamiltonian operators. The advantage of Euler summation is that it
allows us to use term-by-term differentiation on simple-looking but divergent
series. Thus, Euler summation (and more generally other techniques such as Borel
summation and Pad\'e summation) can be used to make sense of the divergent
series that physicists often encounter in their work.

\vspace{0.2cm} 

\noindent {\bf Appendix}. Here we show that the function $K(x,y,t)$ defined in
(\ref{E8}) satisfies the conditions of a delta-convergent series. For fixed $x$
and $0\leq a<b\leq\pi$ we evaluate the integral 
\begin{eqnarray}
&& \int_a^b K(x,y,t)\,{\rd}y\nonumber\\ 
&=& -\sum_{n=1}^{\infty}\left.\frac{t^{n}}{n\pi}[\sin n(x-y)+\sin n(x+y)]
\right|_{y=a}^{y=b}\nonumber\\ 
&=&-2\sum_{n=1}^\infty\frac{t^n}{n\pi}\sin(nx)\,[\cos(nb)-\cos(na)]\nonumber\\
&=&-\frac{{\rm i}}{2\pi}\log\frac{f(x+b)\,f(x-b)\,f(a-x)\,f(-x-a)}{{\bar f}
(x+b)\,{\bar f}(x-b)\,{\bar f}(a-x)\,{\bar f}(-x-a)},\nonumber
\end{eqnarray}
where $f(u)=1-t\,{\re}^{{\rm i}u}$ for $u\in(-\pi,\pi)$, and ${\bar f}$ denotes 
the complex conjugate of the function $f$. Note that 
$$\log\frac{f(u)}{{\bar f}(u)}=2{\rm i}\arg f(u)=2{\rm i}\,\arctan 
\left(\frac{t\,\sin u}{t\,\cos u-1}\right).$$
In the limit $t\to1^-$, $\arg f(u)$ can be found by geometric or trigonometric
methods. In either case we deduce that  
$$\arg f(u)\to\begin{cases} \frac{1}{2}(u-\pi), & u>0\\ 0, & u=0\\
\frac{1}{2}(u+\pi), & u<0\end{cases}.$$
Thus,
\begin{eqnarray}
&& \lim_{t\to1^-}\int_{a}^{b}K(x,y,t)\,{\rm d}y\nonumber\\
&=& \frac{1}{\pi}\left[\frac{1}{2}(x+b-\pi)+\arg f(x-b)\right.\nonumber\\
&& \left.\qquad +\arg f(a-x)+\frac{1}{2}(\pi-x-a)\right]\nonumber\\
&=& \frac{1}{2\pi}(b-a)+\frac{1}{\pi}\left[\arg f(x-b)+\arg f(a-x)\right].
\nonumber
\end{eqnarray}
We conclude that if $x<a$, then 
\begin{eqnarray}
&& \arg f(x-b)+\arg f(a-x)\nonumber\\
&& \qquad=\frac{1}{2}(x-b+\pi)+\frac{1}{2}(a-x-\pi)=\frac{1}{2}(a-b),\nonumber
\end{eqnarray}
so the integral vanishes. Similarly, if $x>b$, then 
\begin{eqnarray}
&& \arg f(x-b)+\arg f(a-x)\nonumber\\
&& \qquad=\frac{1}{2}(x-b-\pi)+\frac{1}{2}(a+\pi-x)=\frac{1}{2}(a-b),\nonumber
\end{eqnarray}
so the integral again vanishes. However, if $a<x<b$, we find that
\begin{eqnarray}
&& \arg f(x-b)+\arg f(a-x)\nonumber\\
&& \qquad=\frac{1}{2}(x-b+\pi)+\frac{1}{2}(a+\pi-x)\nonumber\\
&& \qquad =\frac{1}{2}(a-b+2\pi),\nonumber
\end{eqnarray}
from which we deduce that the value of the integral is unity. This establishes
that $K(x,y,t)$ is indeed a delta-convergent series.

\begin{acknowledgments}
CMB thanks the von Humboldt Foundation for partial financial support.
\end{acknowledgments}

\end{document}